\documentclass[10pt]{amsart}

\usepackage{graphicx}
\usepackage{booktabs}
\usepackage{amsmath}
\usepackage{subcaption}

\usepackage{amsfonts}
\usepackage{amssymb}
\usepackage{amsmath}
\usepackage{amscd}
\usepackage{graphicx}
\usepackage{booktabs}
\usepackage{xcolor}
\usepackage{fancyhdr}
\usepackage{algorithm}
\usepackage{algpseudocode}
\usepackage{float}

\usepackage{mathrsfs}
\usepackage{stmaryrd}

\usepackage[colorlinks,linkcolor=blue,citecolor=blue,urlcolor=blue]{hyperref}
\usepackage[nocompress,space]{cite}

\newcommand{\beq}{\begin{equation}}
\newcommand{\eeq}{\end{equation}}
\newcommand{\bea}{\begin{eqnarray}}
\newcommand{\eea}{\end{eqnarray}}

\definecolor{darkblue}{rgb}{0.0,0.0,0.55}

\newtheorem{definition}{Definition}[section]
\newtheorem{proposition}{Proposition}[section]
\newtheorem{theorem}{Theorem}[section]

\newtheorem{lemma}{Lemma}[section]
\theoremstyle{definition}

\newtheorem{remark}{\textbf{Remark}}[section]

\theoremstyle{definition}	 

\numberwithin{equation}{section}   

\usepackage{lastpage}

\usepackage{mathtools}
\usepackage{mathrsfs}
\usepackage{enumitem}
\usepackage{stmaryrd}
\usepackage{empheq,etoolbox}
\usepackage{xcolor}

\def\BibTeX{{\rm B\kern-.05em{\sc i\kern-.025em b}\kern-.08em
    T\kern-.1667em\lower.7ex\hbox{E}\kern-.125emX}}

\usepackage{bm}
\usepackage{textcomp}
\usepackage{empheq}
\usepackage{booktabs}

\def\Rset{\mathbb{R}}
\def\pdf{f}
\def\gauss{g}
\def\sign{\operatorname{sign}}
\def\esssup{\operatorname{esssup}}

\newlength{\DIFlen}

\begin{document}

\title[Sharp informational inequalities for $\Phi$-entropies]{Sharp moment-entropy and Stam inequalities for $\Phi$-entropies, with applications to the Tsallis, Kaniadakis and general group entropies}

\author{Razvan Gabriel Iagar}
\address{Departamento de Matem\'atica Aplicada, Ciencia e Ingenier\'ia de los Materiales\\
  y Tecnolog\'ia Electr\'onica, Universidad Rey Juan Carlos\\
  28933 M\'ostoles (Madrid), Spain}
\email{razvan.iagar@urjc.es}

\author{David Puertas-Centeno}
\address{Departamento de Matem\'atica Aplicada, Ciencia e Ingenier\'ia de los Materiales\\
  y Tecnolog\'ia Electr\'onica, Universidad Rey Juan Carlos\\
  28933 M\'ostoles (Madrid), Spain \\ and Data, Complex Networks and Cybersecurity\\
  Research Institute, Universidad Rey Juan Carlos, 28028 (Madrid), Spain}
\email{david.puertas@urjc.es}

\author{Piergiulio Tempesta}
\address{Departamento de F\'{\i}sica Te\'{o}rica, Facultad de F\'{\i}sicas\\
  Universidad Complutense de Madrid, 28040 -- Madrid, Spain \\ and Instituto de Ciencias
  Matem\'aticas \\ C/ Nicol\'as Cabrera, No 13--15, 28049 Madrid, Spain}
\email{p.tempesta@fis.ucm.es}


\date{November 23, 2026}

\maketitle

\begin{abstract}
Sharp inequalities for a wide class of functionals known as $\Phi$-entropies, with specific applications for the Tsallis and Kaniadakis entropies and for general group entropies, are established. Employing a general transformation which applies to $\Phi$-entropies, we establish extensions of the triparametric Stam inequality and of the moment-entropy inequality to these classes of entropies. In the case of group entropies, we show how the inequalities can be expressed in terms of the group law. A remarkable property of scaling invariance in the case of the Tsallis entropy is also established. Since generalized entropies are widely used to model complex systems with heavy-tailed statistics, the optimal constants and the explicit extremal densities obtained here provide the quantitative bounds required in maximum-entropy modelling and in the information-theoretic analysis of signals and physical systems.
\end{abstract}

\tableofcontents

\section{Introduction}

The aim of this article is to study a large class of entropic information measures and to establish a number of sharp inequalities satisfied by them. Applications to the most recognized representatives of these generalized entropies are also derived. 

The establishment of sharp inequalities between functionals is recognized as a fundamental component of mathematical analysis, applied mathematics and mathematical physics. Classical functional inequalities, such as the Jensen, H\"{o}lder, Sobolev and Gagliardo--Nirenberg inequalities, have been applied and also generalized in a wide spectrum of scientific disciplines, contributing both to the development of mathematical theories and to the interpretation of physical phenomena. 

The availability of sharp inequalities is also essential in information theory. The relationship between the functionals occupying a prominent position in this field, such as Shannon and Rényi entropies, absolute moments, and Fisher information, are described by several fundamental inequalities that have become indispensable tools in their analysis. In particular, the Stam inequality, originally formulated in \cite{Stam1959}, the moment--entropy inequality, (see for example \cite{Wehrl1978}) and the Cramér--Rao inequality, derived independently in \cite{Rao1945, Cramer1946}, constitute three central results in this area. Let us also mention here that the Cramér-Rao inequality is strongly connected to the central limit theorem, according to \cite{Brown1982}, while more general informational inequalities and connections of them to the Fokker-Planck equation have been given in \cite{Toscani2016(a), Toscani2016(b)} (see also references therein). Complete and detailed discussions of these classical results, together with several generalizations of them, can be found in \cite{Lutwak2004, Lutwak2005, Bercher2012}. The relevance of optimal inequalities for information-theoretic quantities explains also the growing interest in this research direction, as evidenced by many contributions in the latest years (see, for example, \cite{Rioul2010, Ho2015, Sason2016, Yamano2021, Zozor2021, Wu2025}, to name but a few).

Beyond their theoretical interest, these inequalities are basic tools in the applied sciences. The Cram\'er--Rao inequality gives the best accuracy attainable in the estimation of a parameter, and is thus the standard benchmark in statistics and signal processing \cite{Cover2006}. The moment--entropy inequality can be regarded as the sharp form of the maximum entropy principle \cite{Jaynes1957, Kapur1989}: it identifies the least informative density among those with a prescribed moment, which is the usual way of building a probabilistic model from partial data. The Stam inequality is equivalent to the logarithmic Sobolev inequality of Gross \cite{Carlen1991} and is closely related to the uncertainty relations of quantum mechanics \cite{Dembo1991}, whose entropic formulations are given in \cite{Bialynicki1975, Zozor2008}. It also delimits the admissible region of the Fisher--Shannon information plane, which is employed as a diagnostic tool in the analysis of signals and time series \cite{Vignat2003, Contreras-Reyes2023} and of atomic and molecular systems \cite{Romera2004, Esquivel2010, Rong2020}. Finally, the optimal constants of these inequalities are the minimal values of the Cram\'er--Rao and Fisher--Shannon measures of statistical complexity \cite{Rudnicki2016, Zozor2017, Dehesa2023}. In all these applications, it is the knowledge of the optimal constant and of the extremal densities that turns a qualitative principle into a quantitative bound.

Recently, two of the authors and their collaborators have developed further extensions of these inequalities, both by introducing alternative information-theoretic functionals \cite{Zozor2017, Puertas2025} and by employing suitable Sundman-type transformations in order to generalize previous inequalities or enhance their range of admissible parameters \cite{IP2025, IP2025(b), IP2026b}. In particular, the framework introduced in \cite{IP2025(b)} has enabled the derivation of sharp upper estimates for Stam-type and moment--entropy-type products, with the resulting bounds depending explicitly on the regularity properties imposed to the probability density. Sharp informational inequalities have also been established in the relative information-theoretic setting \cite{IPT2025, IP2026}, connecting the Kullback-Leibler and Rényi divergences to new functionals such as a novel scaling-invariant Fisher information.

The notion of entropy has been extended in different directions. A very relevant generalization consists of the \emph{generalized entropies}, whose definition is based on the first three Shannon-Khinchin axioms \cite{Shannon1948, Khinchin1957}. Since the strong additivity axiom is satisfied only by the Shannon (or Boltzmann-Gibbs-Shannon) entropy, a wide class of very interesting generalized entropies is obtained by replacing this additivity axiom by a composability axiom. The latter means that the entropy of a joint distribution is related to the separate entropies of each part by a continuous and symmetric function on $\Rset^2$. We shall make precise these axioms and constructions in the paper, but let us comment here on the (both theoretical and physical) importance of the generalized entropies.

Several well-established entropy functionals have been derived as described above: let us mention here the Rényi entropy (which is a natural generalization of the Shannon one), the Tsallis entropy introduced in \cite{Tsallis1988} and the Kaniadakis entropy defined and studied in \cite{Kaniadakis2001, Kaniadakis2001b, Kaniadakis2002}, among other entropies. An important result established by Enciso and Tempesta in \cite{Enciso2017} proves that the Tsallis entropy (which includes the Shannon one as a limiting case) is the only generalized composable entropy which is in trace form (that is, the entropy is a sum of contributions from individual probabilities). More uniqueness properties of the Tsallis entropy are established in \cite{Tsallis2023b}, while interesting mathematical properties of the Tsallis entropy, including connections to nonlinear diffusion problems, are investigated in \cite{Plastino2022, Suguro2025}. Further applications of the Tsallis entropy to complex systems \cite{Tsallis2023E} and to cosmology \cite{DAgostino2024} have been found. 

Leaving aside the trace-form property, the Kaniadakis entropy proved itself as a very important generalized entropy, motivated by relativistic considerations. Its power-law asymptotic behavior makes it very useful for describing systems exhibiting heavy-tailed distributions, such as high-energy particle spectra and certain complex systems. The Kaniadakis entropy (see for example the theoretical works \cite{Hristopulos2025, Kaniadakis2024} for its foundations) is particularly very fashionable in the last years, having been applied in quantum information \cite{EK2025}, gravity and cosmology \cite{Lymperis2021, Luciano2022} or black-hole thermodynamics \cite{Kaczmarek2024}.

A broader mathematical family of generalized entropies, containing the Tsallis and Kaniadakis entropies as particular cases, is composed by the \emph{group entropies}, whose axiomatic formulation has been given in the foundational papers \cite{Tempesta2011, Tempesta2016a, Tempesta2016b}, where a connection between formal groups and generalized entropies is established. An information-theoretical interpretation of group entropies is given in \cite{Sicuro2016}, where Einstein's likelihood principle is seen as a natural property within the framework of composable group entropies. Other noticeable contributions to the theory of group entropies are given by Jensen and Tempesta in \cite{JensenTempesta2018, JensenTempesta2024}, in the latter the property of extensivity in the thermodynamic limit being employed in order to classify the group entropies depending on the asymptotic behavior of the state space growth rate\cite{Tempesta2020b,Tempesta2020}. Connections of the theory of group entropies with the information geometry are analyzed in \cite{RRT2019}.

A number of sharp informational inequalities satisfied by a very general class of functionals named $\Phi$-entropies, containing as particular cases the Tsallis, Kaniadakis and the whole class of group entropies discussed above, are established in this paper. Besides their wide generality, these inequalities are then particularized to the above mentioned significant representatives of the $\Phi$-entropies. The behavior of the functionals involved in these optimal inequalities with respect to a scaling change is also studied. A remarkable achievement in this study is that, in the case of the Tsallis entropy, it is proved that the scaling parameter plays the role of changing the entropic parameters in the inequality. Such a precise property is no longer valid in the general case, but we instead show how, in the case of the group entropies, the inequalities are expressed in terms of the group law. 

The motivation for extending these inequalities comes from the growing use of generalized entropies as modelling tools. The Tsallis entropy is widely employed in the statistical description of complex systems, from the natural to the social sciences \cite{Tsallis2009b, Tsallis2023E}, including space plasmas \cite{Livadiotis2013}; the Kaniadakis entropy is well suited to phenomena with power-law tails, for instance in plasma physics \cite{Kaniadakis2009, Hristopulos2025}; and generalized entropies of this kind arise as Lyapunov functionals of nonlinear kinetic equations of Fokker--Planck type \cite{Kaniadakis2001, Kaniadakis2001b}. Group entropies, apart from their theoretical role, have been proposed as valuable measures of complexity \cite{Tempesta2020b} satisfying the Shore--Johnson axioms \cite{Jizba2019} (in addition to the first three Shannon--Khinchin axioms), and are potentially relevant in machine learning \cite{CichockiTempesta2026}. Recently, their physical relevance, including black-hole thermodynamics, has been elucidated \cite{JensenJizbaTempesta2026}. In all these contexts one needs, for a generalized entropy, what the classical inequalities provide for the Shannon entropy: the density of maximal entropy under a constraint on a moment, a lower bound for this moment in terms of a Fisher-type information, and the region of the associated information plane that can actually be attained. Under the assumptions of Theorem \ref{th:phi_ent}, the $\Phi$-moment--entropy inequality \eqref{ineq:phi-M-E}, the Cram\'er--Rao-type inequality \eqref{ineq:phi-CR} and the $\Phi$-Stam inequality \eqref{ineq:phi-Stam} provide these three ingredients, with optimal constants and, whenever these are attained, explicit extremal densities.

In order to establish these sharp inequalities, we employ a technique developed by the first two authors in several previous papers (see for example \cite{Puertas2019, IP2025, IP2026, IP2026b}). This technique consists in introducing a one-parameter family of transformations, depending on $k\in\Rset\setminus\{0\}$, together with an exponential-based transformation which is formally the limit of the previous family as $k\to\infty$. These transformations are Sundman-like and measure-preserving ones, in the sense that they involve an implicit change of the independent variable which depends on the own density function, whose aim is to preserve the measure on any differential interval. We then introduce cumulative moments and a Fisher information functional associated to the $\Phi$-entropy, having natural forms in view of the analogy with the standard functionals (to which they reduce in the already well-studied case of the Rényi entropy). Finally, with the aid of these functionals, we transport by our transformations the moment-entropy and Stam inequalities to the general $\Phi$-entropy framework and we derive sharp inequalities for these functionals, as well as their optimizers in view of the inverse transformation. 

For the sake of completeness and easiness of the reading, a number of preliminary notions, functionals, and results are recalled to the reader in Section \ref{sec:prelim}. This is followed by  Section \ref{sec:group-axioms}, where the formal construction of the group entropies is introduced, recalling the Shannon-Khinchin axioms and the composability property, as well as the group logarithms and exponentials. With these preparations, the $\Phi$-transformations of exponential and power type are defined in Section \ref{sec:ineq}, and the inequalities are established by combining the moment-entropy and the generalized (triparametric) Stam inequalities with these transformations. Applications to the Tsallis, Kaniadakis and general group entropies are derived in Section \ref{sec:applications}, which is the final part of the paper.

\section{Preliminary facts}\label{sec:prelim}

We recall in this section for the sake of completeness the main informational functionals employed throughout the paper, together with the moment-entropy and Stam inequalities. Whenever is employed, the notation $p^*=\frac{p}{p-1}$ designs the H\"older conjugate of $p$. 

\medskip

\noindent \subsection{The $p$-th absolute moment.} The $p$-th absolute moment of a probability density function $f$, with $p \geqslant 0$, is given by
\begin{equation}\label{eq:pabs}
\mu_p[\pdf] = \left\langle  \, |x|^p \, \right\rangle_\pdf:= \int_\Rset |x|^p \, \pdf(x) \,dx.
\end{equation}
Related to the $p$-th absolute moment, it is more common to consider the related quantity known as the $p$-th deviation, defined by taking the power $1/p$ in~\eqref{eq:pabs} as follows:
\begin{equation*}
\sigma_p[\pdf] =\left(\int_\Rset |x|^p \, \pdf(x) \,dx \right)^{\frac{1}{p}}, \quad {\rm for} \ p > 0.
\end{equation*}
{We can also define deviations in the limiting cases $p\to0$ and $p\to\infty$ as indicated below.}
\begin{equation*}
\begin{split}
&\sigma_0[\pdf] = \lim\limits_{p \to 0} \sigma_p[\pdf]  = \exp\left(\int_\Rset \pdf(x) \, \log|x| \,dx \right).
\\
&\sigma_{\infty}[\pdf]=\lim\limits_{p\to\infty}\sigma_p[\pdf]=\esssup\big\{|x| : x\in\Rset, \pdf(x) > 0 \big\},
\end{split}
\end{equation*}

\medskip

\noindent \subsection{Shannon, Rényi and Tsallis entropies.} 

We {recall here the definitions of the two classical entropy functionals.}
\begin{definition} 
The Shannon entropy of a probability density function $\pdf$ is defined as
\begin{equation*}
	S[\pdf]=-\int_\Rset \pdf(x) \, \log\pdf(x) \,dx.
\end{equation*}
\end{definition}
\begin{definition}
The Rényi and Tsallis entropies of $\lambda-$order, $\lambda\neq1$, of a probability density function $\pdf$ defined on $\Rset$ (or on a Lebesgue measurable subset of $\Rset$)  are defined by
\begin{equation*}
R_\lambda[\pdf] = \frac1{1-\lambda} \log\left( \int_\Rset [\pdf(x)]^\lambda \,dx \right), \quad {\rm and} \quad T_\lambda[\pdf] = \frac1{\lambda-1} \left( 1 - \int_\Rset [\pdf(x)]^\lambda \,dx \right),
\end{equation*}
respectively. 
\end{definition}
{One can easily obserye that the Rényi entropies} are one-parameter generalizations of the Shannon entropy, {since the latter is recovered as the limiting case $\lambda\to1$. Another simple remark is that} the Rényi and Tsallis entropies are one-to-one mapped:
$$
T_{\lambda}[\pdf]=\frac{e^{(1-\lambda) R_{\lambda}[\pdf]} - 1}{1 - \lambda}.
$$
Throughout this paper, we employ the quantity known as the Rényi entropy power, that is,
$$
N_\lambda[\pdf] = e^{R_{\lambda}[\pdf]} = \left\langle\pdf^{\lambda-1}(x)\right\rangle^\frac1{1-\lambda}_\pdf.
$$
We also denote by $N[\pdf]=e^{S[\pdf]}$ the Shannon entropy power.

\medskip

\noindent \subsection{$(p,\lambda)-$Fisher information.} The $(p,\lambda)-$Fisher information is a generalization of the usual (non-parametric) Fisher information
\begin{equation*}
	F[f]=\int_\Rset f(x) \left(\frac{f' (x)}{f(x)}\right)^2\,dx
\end{equation*}
and it was introduced by Lutwak and Bercher~\cite{Lutwak2005, Bercher2012, Bercher2012a}. 
\begin{definition}
Given $p>0$ and $\lambda \in \Rset\setminus\{0\}$, the $(p,\lambda)-$Fisher information of a probability density function $f$ which is assumed to be derivable on its support is defined as
\begin{equation}\label{eq:def_FI}
F_{p,\lambda}[\pdf]\: = \: \int_\Rset \left|\pdf(x)^{\lambda-2} \, \frac{{\rm d}\pdf}{{\rm d}x}(x)\right|^{p} \pdf(x) \,dx.
\end{equation}
\end{definition}
The following notation is also frequently employed throughout the paper:
\begin{equation}
\phi_{p,\lambda}[f]=\left(F_{p,\lambda}[f]\right)^\frac{1}{p\lambda}.
\end{equation}
Note that $F[f]=F_{2,1}[f]$, thus the non-parametric Fisher information corresponds to $(p,\lambda)=(2,1)$. Another case of interest is the $(1,\lambda)-$Fisher information, which corresponds to the total variation of $\frac{\pdf^\lambda}{\lambda}$ whenever $\pdf^\lambda$ is a function with bounded variation.

\medskip

\noindent \subsection{The moment-entropy inequality} It is an informational inequality relating the R\'enyi power entropy and the moments $\sigma_p$. More precisely, when
\begin{equation}\label{eq:param_clas}
p^*\in[0,\infty), \quad {\rm and} \quad \lambda>\frac1{1+p^*},
\end{equation}
it was proved in~\cite{Lutwak2004, Lutwak2005, Bercher2012} that, for any probability density function $f$, we have
\begin{equation}\label{ineq:bip_E-M}
\frac{\sigma_{p^*}[\pdf]}{N_{\lambda}[\pdf]} \, \geqslant \, \frac{\sigma_ {p^*}[\gauss_{p,\lambda}]}{N_{\lambda}[\gauss_{p,\lambda}]}\equiv K^{(0)}_{p,\lambda},
\end{equation}
and the minimizers of Eq. \eqref{ineq:bip_E-M} are the functions $g_{p,\lambda}$ known as generalized Gaussians \cite{Lutwak2005}, $q-$Gaussians \cite{Bercher2012a} or stretched deformed Gaussians \cite{Zozor2017}. For $p>1$, these minimizers are given by
\begin{equation}\label{def:g_plambda}
\gauss_{p,\lambda}(x) \, = \, \frac{a_{p,\lambda}}{\exp_\lambda\left( |x|^{p^*} \right)}
\, = \, a_{p,\lambda}\, \exp_{2-\lambda}\left( - |x|^{p^*} \right),
\end{equation}
where $\exp_\lambda$ is the generalized Tsallis exponential
\begin{equation}\label{def:q-exp}
\exp_\lambda(x):= \left( 1 + (1 - \lambda) \, x \right)_+^\frac1{1-\lambda}, \ \ \lambda \ne 1, \quad \exp_1(x)=\exp(x),
\end{equation}
and $a_{p,\lambda}$ has an explicit value which we omit here (see \cite{IP2025}).

\medskip

\noindent \subsection{\texorpdfstring{Differential-escort transformations}{Differential-escort transformations}}\label{sec:escort} The following transformations, introduced in \cite{Puertas2019}, are used throughout the paper.
\begin{definition}\label{def:escort}
Let $f$ be a probability density function, positive on an open interval and zero outside it, let $y_0$ be a point of this interval and let $\theta\in\Rset\setminus\{0\}$. The \emph{differential-escort transformation} of exponent $\theta$ of $f$ is
\begin{equation}\label{eq:escort}
\mathfrak{E}_{\theta}[f](z):=f(y(z))^{\theta},\qquad z(y)=\int_{y_0}^{y}f(t)^{1-\theta}\,dt,
\end{equation}%
extended by zero outside the image of the support of $f$.
\end{definition}
Observe that, since $\mathfrak{E}_{\theta}[f]\,dz=f\,dy$, $\mathfrak{E}_{\theta}[f]$ is again a probability density function, and a different choice of $y_0$ only produces a translation. {Moreover,}
\[
{\frac{S\big[\mathfrak{E}_{\theta}[f]\big]}{S[f]}=\frac{R_{\lambda}\big[\mathfrak{E}_{\theta}[f]\big]}{R_{1+\theta(\lambda-1)}[f]}=\frac{T_{\lambda}\big[\mathfrak{E}_{\theta}[f]\big]}{T_{1+\theta(\lambda-1)}[f]}=\theta},\quad  \mu_{p}\big[\mathfrak{E}_{\theta}[f]\big]=\int_{\Rset}|z(y)|^{p}f(y)\,dy,
\]
the latter being the $p$-th moment of $\mathfrak{E}_{\theta}[f]$ centered at $z(y_0)=0$.

\medskip

\noindent \subsection{The extended (triparametric) Stam inequality} This is a generalized inequality, valid both in the classical domain of parameters and in a mirrored domain, established first in \cite{Zozor2017} and then extended by the authors in the previous work \cite[Theorem 5.1]{IP2025} with the help of the {newly defined} up and down transformations. Let $p\geqslant 1$ and $q$ be such that
\begin{equation}\label{eq:sign_cond1}
\sign\left(p^*q + \lambda-1\right)=\sign\left(q+1-\lambda\right)\neq0.
\end{equation}
Then, the following generalized Stam inequality holds true for any probability density function $f$ such that $\mathfrak{E}_{1+q-\lambda}[f]$ is absolutely continuous on $\Rset$:
\begin{equation}\label{ineq:trip_Stam_extended}
	\left(\phi_{p,q}[\pdf] \, N_{\lambda}[\pdf] \right)^{\theta(q,\lambda)}\: \geqslant \:	\left(\phi_{p,q}[\widetilde g_{p,q,\lambda}] \, N_{\lambda}[\widetilde g_{p,q,\lambda}] \right)^{\theta(q,\lambda)}\equiv \kappa_{p,q,\lambda}^{(1)},
\end{equation}
where $\theta(q,\lambda)=1+q-\lambda$ and $\widetilde g_{p,q,\lambda}$ is defined in \cite[Section 5]{IP2025}. Moreover, for $p<1$ and $q, \lambda$ such that
\begin{equation}\label{eq:sign_cond2}
\sign(p^*q+\lambda-1)=\sign(q+\lambda-1)\neq0, \quad \sign(\lambda-1)=\sign(q)\neq0.
\end{equation}
the inequality ~\eqref{ineq:trip_Stam_extended} holds true with $\theta(q,\lambda)=-q$, provided that $f:(x_i,x_f)\mapsto\Rset^+$, $x_i>-\infty$, is continuously differentiable with $f'<0$. The optimal constants and the minimizers $\widetilde g_{p,q,\lambda}$ are known, but we omit the (rather technical) expressions here and we refer the interested reader to \cite[Section 5]{IP2025} for them.

\section{Group entropies}\label{sec:group-axioms}

This section is devoted to a detailed review of the general notion of group entropy. Group entropies form a broad class of generalized entropies, first introduced and discussed in \cite{Tempesta2011}, and subsequently extended in a series of works (see, for instance, \cite{Tempesta2016a, Tempesta2016b, RRT2019, Tempesta2020, Tempesta2020b}, and references therein).

\subsection{Beyond the Shannon-Khinchin formulation: the Composability Axiom}\label{sec:21}

\begingroup\sloppy
Let $\{p_{i}\}_{i=1,\cdots ,W}$, with $W\geq 2$, be a discrete probability distribution, i.e., $0\leq p_{i}\leq 1$ and $\sum_{i=1}^{W} p_{i}=1$. We denote by $\mathcal{P}$ the set of all discrete probability distributions with $W$ components, for every $W\geq 2$. Let $S$ be a positive function defined on $\mathcal{P}$. The first three \textit{Shannon-Khinchin} (SK) axioms \cite{Shannon1948, Khinchin1957} may be summarized as follows: \newline

\noindent \textbf{(SK1)} \textit{Continuity}: $S(p)$ is continuous with respect to all variables $p_{1},\ldots,p_{W}$. \newline
\noindent \textbf{(SK2)} \textit{Maximality}: $S(p)$ achieves its maximum at the uniform distribution. \newline
\noindent \textbf{(SK3)} \textit{Expansibility}: adding an event with zero probability does not affect the value of $S(p)$.

A function satisfying (SK1)--(SK3) is called a \textit{generalized entropy}.

Supplemented by the strong additivity, these axioms characterize uniquely the \textit{Boltzmann--Gibbs--Shannon} (BGS) entropy
\begin{equation}
S_{BGS}(p)=S_{BGS}(p_{1},p_{2},...,p_{W})=-\sum_{i=1}^{W}p_{i}\ln p_{i}
\label{S_BGS}
\end{equation}
up to a positive multiplicative factor.

In terms of two random variables $X$ and $Y$, the strong additivity reads
\begin{equation}
S(X,Y)=S(X)+S(Y\left\vert X\right),
\label{separability}
\end{equation}
where $S(X,Y)$ is the entropy of the joint distribution $p(x,y)$ and the \textit{conditional entropy} $S(Y\left\vert X\right)$ is the average, with respect to $X$, of the entropies $S(Y\left\vert X=x\right)$.

The idea behind group entropies \cite{Tempesta2016a, Tempesta2016b} is to replace the strong additivity \eqref{separability} by the following \textit{composability axiom}, which connects entropies with formal group theory (for the role of group entropies in information geometry, see \cite{RRT2019}).

\begin{definition}[Composability Axiom]
\label{composab} \label{Def1}We say that a generalized entropy $S$ is
\textit{composable} if there exists a continuous function of two real
variables $\Phi (x,y)$, called the composition law, such that the following
four conditions are satisfied.

\begin{description}
\item[(C1)] Composability: $S(p\times q)=\Phi (S(p),S(q))$ for any $p,q\in
\mathcal{P}.$ \label{con:com}

\item[(C2)] Symmetry: $\Phi (x,y)=\Phi (y,x)$. \label{con:symm}

\item[(C3)] Associativity: $\Phi (x,\Phi (y,z))=\Phi (\Phi (x,y),z)$. \label%
{con:asoc}

\item[(C4)] Null-composability: $\Phi (x,0)=x$. \label{con:null}
\end{description}
\end{definition}


Conditions (C2)--(C4) make the composition law physically admissible: the entropy of a composite system does not depend on the labelling of its subsystems (C2), nor on the order in which they are combined (C3), a property that is crucial for a consistent zeroth law of thermodynamics, and composing with a system of zero entropy leaves the entropy unchanged (C4).

\begin{definition}
\label{def:groupentropy} A group entropy is a function $S:\mathcal{P}%
\rightarrow \mathbb{R}^{+}\cup \{0\}$ which satisfies the Shannon-Khinchin
axioms (SK1)-(SK3) and the composability axiom (C1)-(C4).
\end{definition}
\begin{definition}
If an entropy satisfies the SK axioms (SK1)-(SK3) and the composability axiom only on the uniform distribution, we shall say that it is a weak group entropy.
\end{definition}
\begin{remark}
As shown by a theorem established in \cite{Enciso2017}, entropy functions of trace form are restricted to be weak group entropies, with the exception of the Tsallis entropy, which constitutes a genuine example of group entropy.
\end{remark}

\endgroup
The construction of group entropies that we present is motivated by the concept of a formal group law, which we introduce in the following.

\subsection{Formal group laws}\label{sec:22}

In the construction that follows, all rings under consideration are assumed to be associative, commutative, and unital. In particular, for a ring $(A,+,\cdot)$, we denote by $0$ the identity element of the additive operation $+:A\times A\rightarrow A$, and by $1$ the identity element of the multiplicative operation $\cdot :A\times A\rightarrow A$.
\begin{definition}
Let $R$ be a commutative ring with identity, and $R\left\{
x_{1},x_{2},..\right\} $ be the ring of formal power series in the variables
$x_{1}$, $x_{2}$, ... with coefficients in $R$. A commutative
one-dimensional formal group law over $R$ is a formal power series in two
variables $\Psi \left( x,y\right) \in R\left\{ x,y\right\} $ of the form $%
\Psi \left( x,y\right) =x+y+$ \emph{terms of higher degree}, such that
\begin{equation*}
i)\qquad \Psi \left( x,0\right) =\Psi \left( 0,x\right) =x
\end{equation*}%
\begin{equation*}
ii)\qquad \Psi \left( \Psi \left( x,y\right) ,z\right) =\Psi \left( x,\Psi
\left( y,z\right) \right) \text{.}
\end{equation*}
When $\Psi \left( x,y\right) =\Psi \left( y,x\right)$, the formal group law
is said to be commutative.
\end{definition}
For any formal group law $\Psi (x,t)$, there exists an \textit{inverse
formal series} $\psi \left( x\right) $ $\in R\left\{ x\right\} $ such
that $\Psi \left( x,\psi \left( x\right) \right) =0$. This is the reason
why we talk about \textit{formal group laws}. The interested reader might see \cite{Hazewinkel1978, Serre1992} for a deeper exposition on formal group theory.

Some relevant examples are listed below.

\begin{itemize}
\item The \textit{additive} \textit{group law}
\begin{equation*}
\Psi (x,y)=x+y\ .
\end{equation*}

\item The \textit{multiplicative} \textit{group law}
\begin{equation*}
\Psi (x,y)=x+y+axy\ .
\end{equation*}

\item The \textit{hyperbolic} \textit{group law} (addition of velocities in
special relativity)
\begin{equation*}
\Psi (x,y)=\frac{x+y}{1+xy}\ .
\end{equation*}

\item The \textit{Euler group law}
\begin{equation*}
\Psi (x,y)=(x\sqrt{1-y^{4}}+y\sqrt{1-x^{4}})/(1+x^{2}y^{2})\ .
\end{equation*}%
which appears in the sum of elliptic integrals
\begin{equation*}
\int_{0}^{x}\frac{dt}{\sqrt{1-t^{4}}}+\int_{0}^{y}\frac{dt}{\sqrt{1-t^{4}}}%
=\int_{0}^{\Psi (x,y)}\frac{dt}{\sqrt{1-t^{4}}}\ .
\end{equation*}
\end{itemize}

In the construction of a group entropy, the composition rule can be described algebraically as a commutative formal group law, namely a continuous function of the form
\begin{equation}
\Psi (x,y)=\sum_{n=0}^{\infty }\sum_{i=0}^{n}a_{i,n-i}x^{i}y^{n-i}
\label{Phi(x,y)}
\end{equation}
which is required to satisfy the conditions (C2)–(C4) of Definition \ref{Def1}. The symmetry condition then implies that the coefficients obey $a_{i,n-i}=a_{n-i,i}$ for all $n\geq 1$ and $0\leq i\leq n$. Moreover, the null-composability property enforces $a_{0,0}=0$, $a_{1,0}=a_{0,1}=1$, together with $a_{n,0}=0$ for all $n\geq 2$.

This structure underlies the connection between entropic measures and formal group theory, first pointed out in \cite{Tempesta2011}, which will be further illustrated in the subsequent discussion.
\subsection{Group logarithms and exponentials}\label{sec:23}

Our construction is based on the notion of a group logarithm, which can be associated with any formal group law. The precise regularity assumptions involved in its definition may vary depending on the specific application under consideration, reflecting a certain degree of flexibility in its formulation. In particular, the usual logarithm corresponds to the additive formal group law.

\begin{definition}
A group logarithm is a strictly increasing and strictly concave function $%
\log _{G}:(0,\infty )\rightarrow \mathbb{R}$, with $\log _{G}(1)=0$
(possibly depending on a set of real parameters), satisfying a functional
equation of the form
\begin{equation}
\log _{G}(xy)=\chi (\log _{G}(x),\log _{G}(y))  \label{glog}
\end{equation}%
where $\chi (x,y)$ fulfills the requirements $(C2)$-$(C4)$. Equation (\ref%
{glog}) will be called the group law associated with $\log _{G}(\cdot )$.
\end{definition}

In what follows, we restrict our attention to group logarithms of the form
\begin{equation}
\log_{G}(x)=G(\ln x)\ .
\label{eq:GL}
\end{equation}
In this expression, $G$ denotes a strictly increasing function satisfying the local expansion $G(t)=t+O(t^{2})$ as $t\to 0$, which in particular implies $G(0)=0$. The strict concavity of $\log_{G}(x)$ amounts to the additional requirement $G''(t)<G'(t)$, since $\frac{d^{2}}{dx^{2}}G(\ln x)=\big(G''(\ln x)-G'(\ln x)\big)/x^{2}$. For a more detailed discussion of sufficient conditions guaranteeing that $G$ meets these requirements, we refer to \cite{Tempesta2016b}.

A first important example of a nontrivial group logarithm is provided by the so-called $q$-logarithm. In this case one has
\begin{equation*}
G(t)=\frac{\mathrm{e}^{(1-q)t}-1}{1-q},\qquad \log _{q}(x)=G(\ln x)=\frac{x^{1-q}-1}{1-q},\qquad q>0,
\label{qlog}
\end{equation*}
which has been studied in the context of nonextensive statistical mechanics \cite{Tsallis2009b}.

The inverse function (in the sense of functional inversion) of a group logarithm is referred to as the corresponding group exponential, and is defined by
\begin{equation}
\exp_{G}(x)=e^{G^{-1}(x)}.
\label{Gexp}
\end{equation}
In the special case $G(t)=t$, one recovers the standard exponential function. Likewise, when $G(t)=\frac{e^{(1-q)t}-1}{1-q}$ as above, the associated group exponential reduces to the $q$-exponential
\begin{equation*}
e_{q}(x)=\left[1+(1-q)x\right]_{+}^{\frac{1}{1-q}},
\end{equation*}
and analogous constructions generate infinitely many other examples of group logarithms and exponentials, as discussed for instance in \cite{Tempesta2011}.

Generalized logarithms play a central role in the construction of group entropies, as they encode the composition rules through the underlying formal group structure. A prominent example within this framework is given by the class of $Z$-entropies introduced in \cite{Tempesta2016b}. Their general form, for $\alpha>0$, is
\begin{equation*}
Z_{G,\alpha}(p):=\frac{1}{1-\alpha}\log_{G}\Big(\sum_{i=1}^{W}p_{i}^{\alpha }\Big),
\end{equation*}
where $\log_{G}$ denotes a generic group logarithm.

\section{General inequalities for $\Phi$-entropies}\label{sec:ineq}

This section is dedicated to the deduction of sharp informational inequalities for general $\Phi$-entropies. To this end, we introduce a transformation acting on probability density functions and we apply the Stam and moment-entropy inequalities to the transformed densities. We begin by recalling the notion of $\Phi$-entropy, following \cite{Toranzo2018}.

	\begin{definition}[$\Phi$-entropy]\label{def:phi_ent}
		Let $\Phi:[0,\infty)\to\Rset$ be a convex function such that $\Phi(0)=0$ and $\Phi\in C^2((0,\infty))$, and let $f$ be a probability density function. The $\Phi$-entropy of $f$ is defined by
		\begin{equation}\label{eq:phi_ent}
			H_{\Phi}[f]:=-\int_{\Rset}\Phi(f(x))\,dx.
		\end{equation}
	\end{definition}
	
	We observe that this general notion of entropy includes a number of well-established entropies as particular cases. For example
	
	$\bullet$ the classical Shannon entropy, introduced in \cite{Shannon1948}, corresponds to $\Phi(t)=t\log\,t$ for $t>0$ (with $\Phi(0)=0$).
	
	$\bullet$ the Tsallis entropies \cite{Tsallis1988} correspond to the entropic functional
	\begin{equation}\label{eq:Tsallis}
		\Phi(t)=\frac{t^{\alpha}-t}{\alpha-1}, \qquad \alpha>0,\ \alpha\neq1.
	\end{equation}
	
	$\bullet$ the recently very fashionable Kaniadakis entropies, introduced in \cite{Kaniadakis2001, Kaniadakis2002}, correspond to the entropic functional
	\begin{equation}\label{eq:Kania}
		\Phi(t)=\frac{t^{1+\alpha}-t^{1-\alpha}}{2\alpha}, \qquad \alpha\in(-1,1),\ \alpha\neq0.
	\end{equation}
Inspired by the previous remarkable examples, we assume throughout the paper that the function $\Phi(t)/t$ is invertible for $t>0$ and denote by
\begin{equation}\label{eq:varphi}
\varphi^{-1}(t):=\frac{\Phi(t)}{t}.
\end{equation}	
	
\subsection{Exponential based $\Phi$-trasformation}

Let us next define the following transformation, acting on probability density function, that will be called \emph{the $\Phi$-transformation} throughout the paper.

\begin{definition}[$\Phi$-transformation]\label{def:phi_transf}
	Let $\Phi$ and $f$ as in Definition \ref{def:phi_ent}. We define the \emph{$\Phi$-transformation} as follows:
	\begin{equation}\label{eq:phi_transf}
		\mathfrak{Z}_{\Phi}[f](y):=\exp\left(\frac{\Phi(f(x(y)))}{f(x(y))}\right), \qquad y'(x):=f(x)\exp\left(-\frac{\Phi(f(x))}{f(x)}\right).
	\end{equation}
	For simplicity, we will employ the shorter notation
	$$
	\widehat{f_{\Phi}}\equiv\mathfrak{Z}_{\Phi}[f].
	$$
\end{definition}
We remark that the change of variable is defined up to a translation, as $y(x)$ depends on an integration constant. We write throughout the paper
\begin{equation}\label{eq:y(x)}
	y(x)=\int_{x_0}^xf(t)\exp\left(-\frac{\Phi(f(t))}{f(t)}\right)\,dt,
\end{equation}
where $x_0$ is a fixed point in the support of $f$. Throughout {this section}, $f$ is assumed to be positive on an open interval (its support) and zero outside it, and $\widehat{f_{\Phi}}$ is extended by zero outside the image of the support. 
Note also that, if $\Phi(t)=t\log\,t$, corresponding to the Shannon entropy, the transformation $\mathfrak{Z}_{\Phi}$ is just the identity.

We next establish two basic properties of the $\Phi$-transformation.
\begin{lemma}\label{lem:deriv}
	Let $f$ be a probability density function. Then $\widehat{f_{\Phi}}$ is a probability density function as well. Moreover, if $f$ is differentiable, then $\widehat{f_{\Phi}}$ is also differentiable and
	\begin{equation}\label{eq:deriv_phi}
		\frac{d\widehat{f_{\Phi}}(y)}{dy}=\exp\left(\frac{2\Phi(f(x))}{f(x)}\right)\frac{f'(x)}{f(x)^3}\left(\Phi'(f(x))f(x)-\Phi(f(x))\right).
	\end{equation}
\end{lemma}
\begin{proof}
	We observe that
	\begin{equation*}
		\begin{split}
			\int_{\Rset}\widehat{f_{\Phi}}(y)\,dy&=\int_{\Rset}\exp\left(\frac{\Phi(f(x(y)))}{f(x(y))}\right)\,dy=\int_{\Rset}\exp\left(\frac{\Phi(f(x))}{f(x)}\right)y'(x)\,dx\\
			&=\int_{\Rset}\exp\left(\frac{\Phi(f(x))}{f(x)}\right)\exp\left(-\frac{\Phi(f(x))}{f(x)}\right)f(x)\,dx=\int_{\Rset}f(x)\,dx=1,
		\end{split}
	\end{equation*}
	hence $\widehat{f_{\Phi}}$ is a probability density function. Eq. \eqref{eq:deriv_phi} follows from the chain rule by direct calculation as shown below:
	\begin{equation*}
		\begin{split}
			\frac{d\widehat{f_{\Phi}}(y)}{dy}&=\frac{d}{dx}\exp\left(\frac{\Phi(f(x))}{f(x)}\right)x'(y)\\
			&\hspace{-1em}=\exp\left(\frac{\Phi(f(x))}{f(x)}\right)\frac{\Phi'(f(x))f(x)f'(x)-\Phi(f(x))f'(x)}{f^2(x)}\exp\left(\frac{\Phi(f(x))}{f(x)}\right)f(x)^{-1}\\
			&\hspace{-1em}=\exp\left(\frac{2\Phi(f(x))}{f(x)}\right)\frac{f'(x)}{f(x)^3}\left(\Phi'(f(x))f(x)-\Phi(f(x))\right),
		\end{split}
	\end{equation*}
	completing the proof.
\end{proof}

We next introduce the inverse transformation to the $\Phi$-transformation.
\begin{definition}
	The transformation
\begin{equation}\label{eq:inverse}
	\mathfrak{Z}_{\Phi}^{-1}[f](s):=\varphi(\log\,f(x(s))), \quad s(x):=\int_{x_0}^{x}\frac{f(t)}{\varphi(\log\,f(t))}\,dt
\end{equation}	
is the inverse of the $\Phi$-transformation defined in Eq. \eqref{eq:phi_transf}.
\end{definition}
The existence of a well-defined inverse transformation allows us to recover information from the application of the transformation $\mathfrak{Z}_{\Phi}$ and in particular for deriving minimizers for the forthcoming sharp inequalities.

Based on the previous transformations, we introduce generalized $(\Phi,p,\gamma)$-moments and generalized $(\Phi,p,\beta)$-Fisher information functionals by means of the following definitions.
\begin{definition}\label{def:inf_funct}
	Let $\Phi$ and $f$ be as in Definition \ref{def:phi_ent}. Then, for any $p\geq0,\,\gamma\in\Rset$  and $\beta\in\Rset\setminus\{0\}$, we define the $(\Phi,p,\gamma)$-moments of $f$ by
	\begin{equation}\label{eq:phi_moment}
		\mu_{p,\gamma}^{[\Phi]}[f]:=\int_{\Rset}\left|\int_{x_0}^{x}\exp\left(-\gamma\frac{\Phi(f(t))}{f(t)}\right)f(t)\,dt\right|^pf(x)\,dx,
	\end{equation}
	the $(p,\gamma)$-moments of a probability density $g$ with respect to a base point $y_0$ by
	\[
		\mu_{p,\gamma}[g]:=\int_{\Rset}\left|\int_{y_0}^{y}g(t)^{1-\gamma}\,dt\right|^{p}g(y)\,dy,
	\]%
	and the exponential $\Phi$-entropy of order $\lambda$ by
\begin{equation}\label{eq:phi_exp_ent}
	\widehat{N}_{\lambda}^{[\Phi]}[f]:=\left[\int_{\Rset}\exp\left(\frac{(\lambda-1)\Phi(f(x))}{f(x)}\right)f(x)\,dx\right]^{\frac{1}{1-\lambda}}.	
\end{equation}		
Moreover, if $f$ is differentiable, the $(\Phi,p,\beta)$-Fisher information of $f$ is given by
	\begin{equation}\label{eq:phi_Fisher}
		F_{p,\beta}^{[\Phi]}[f]:=\int_{\Rset}f(x)^{1-p}\exp\left(\frac{p\beta\Phi(f(x))}{f(x)}\right)\left|\frac{d}{dx}\left(\frac{\Phi(f)}{f}\right)(x)\right|^p\,dx.
	\end{equation}
	We also define
	$$
	\phi_{p,\beta}^{[\Phi]}[f]=\left[F_{p,\beta}^{[\Phi]}[f]\right]^\frac1{p\beta}.
	$$
\end{definition}

\noindent The following result motivates the definition of the previous informational functionals.
\begin{proposition}\label{prop:funct}
	Let $f$ and $\Phi$ be as in Definition \ref{def:phi_ent}. Then, with $y_0=y(x_0)$,
	\begin{equation}\label{eq:phi_moment2}
		\mu_{p,\gamma}^{[\Phi]}[f]=\mu_{p,\gamma}[\widehat{f_{\Phi}}], \quad p\geq0,
	\end{equation}
	and, for $\gamma\neq0$, with $\mathfrak{E}_{\gamma}$ given by \eqref{eq:escort} with base point $y_0$,
	\begin{equation}\label{eq:phi_moment_escort}
		\mu_{p,\gamma}[\widehat{f_{\Phi}}]=\mu_{p}\big[\mathfrak{E}_{\gamma}[\widehat{f_{\Phi}}]\big], \quad p\geq0,
	\end{equation}%
	and
	\begin{equation}\label{eq:phi_Fisher2}
		F_{p,\beta}^{[\Phi]}[f]=F_{p,\beta}[\widehat{f_{\Phi}}].
	\end{equation}
	Moreover,
\begin{equation}\label{eq:phi_exp_ent2}
	\widehat{N}_{\lambda}^{[\Phi]}[f]=N_{\lambda}[\widehat{f_{\Phi}}]
\end{equation}		
and	the following identity involving the Shannon entropy is remarkable:
	\begin{equation}\label{eq:phi_Renyi}
		S[\widehat{f_{\Phi}}]=H_{\Phi}[f].
	\end{equation}
\end{proposition}
\begin{proof}
	Recalling \eqref{eq:y(x)}, we infer from the definition of $\mu_{p,\gamma}$ that
	\begin{equation*}
		\begin{split}
			\mu_{p,\gamma}[\widehat{f_{\Phi}}]&=\int_{\Rset}\left|\int_{y_0}^y\left[\widehat{f_{\Phi}}(t)\right]^{1-\gamma}\,dt\right|^p\widehat{f_{\Phi}}(y)\,dy\\&=\int_{\Rset}\left|\int_{x_0}^{x}\exp\left(-\gamma\frac{\Phi(f(t))}{f(t)}\right)f(t)\,dt\right|^pf(x)\,dx=\mu_{p,\gamma}^{[\Phi]}[f].
		\end{split}
	\end{equation*}
	The second equality follows from the changes of variables $t=y(s)$ in the inner integral and $y=y(x)$ in the outer one. Indeed, by \eqref{eq:phi_transf} and \eqref{eq:y(x)}, $y$ is strictly increasing on the support of $f$, with
$y(x_0)=y_0$ {and 
$$
\widehat{f_{\Phi}}(y(s))=\exp\left(\frac{\Phi(f(s))}{f(s)}\right),\quad y'(s)=f(s)\exp\left(-\frac{\Phi(f(s))}{f(s)}\right),
$$ 
whence}
	\[
		\big[\widehat{f_{\Phi}}(y(s))\big]^{1-\gamma}y'(s)=f(s)\exp\left(-\gamma\frac{\Phi(f(s))}{f(s)}\right),\qquad \widehat{f_{\Phi}}(y(x))\,y'(x)=f(x).
	\]%
	In order to prove \eqref{eq:phi_moment_escort}, let $\gamma\neq0$. The change of variable $z(y)=\int_{y_0}^{y}\big[\widehat{f_{\Phi}}(t)\big]^{1-\gamma}\,dt$ in \eqref{eq:escort} is strictly increasing on the support of $\widehat{f_{\Phi}}$, with $z(y_0)=0$ and
	\[
		\mathfrak{E}_{\gamma}[\widehat{f_{\Phi}}](z(y))\,z'(y)=\widehat{f_{\Phi}}(y)^{\gamma}\,\widehat{f_{\Phi}}(y)^{1-\gamma}=\widehat{f_{\Phi}}(y).
	\]%
	Therefore
\begin{equation*}
		\begin{split}
			\mu_{p}\big[\mathfrak{E}_{\gamma}[\widehat{f_{\Phi}}]\big]&=\int_{\Rset}|z|^{p}\,\mathfrak{E}_{\gamma}[\widehat{f_{\Phi}}](z)\,dz=\int_{\Rset}|z(y)|^{p}\,\widehat{f_{\Phi}}(y)\,dy\\
			&=\int_{\Rset}\left|\int_{y_0}^{y}\left[\widehat{f_{\Phi}}(t)\right]^{1-\gamma}\,dt\right|^{p}\widehat{f_{\Phi}}(y)\,dy=\mu_{p,\gamma}[\widehat{f_{\Phi}}].
		\end{split}
	\end{equation*}%
In order to prove \eqref{eq:phi_Fisher2}, we take into account \eqref{eq:deriv_phi} and compute
	\begin{equation*}
		\begin{split}
			F_{p,\beta}[\widehat{f_{\Phi}}]&=\int_{\Rset}\widehat{f_{\Phi}}^{1+(\beta-2)p}(y)|\widehat{f_{\Phi}}'(y)|^p\,dy\\
			&\hspace{-2.5em}=\int_{\Rset}f(x)\exp\left(\frac{p(\beta-2)\Phi(f(x))}{f(x)}\right)\left|\exp\left(\frac{2\Phi(f(x))}{f(x)}\right)\frac{1}{f(x)}\frac{d}{dx}\left(\frac{\Phi(f(x))}{f(x)}\right)\right|^p\,dx\\
			&\hspace{-2.5em}=\int_{\Rset}f(x)^{1-p}\exp\left(\frac{p\beta\Phi(f(x))}{f(x)}\right)\left|\frac{d}{dx}\left(\frac{\Phi(f(x))}{f(x)}\right)\right|^p\,dx=F_{p,\beta}^{[\Phi]}[f],
		\end{split}
	\end{equation*}
	as desired. Finally, we calculate
	$$
	N_{\lambda}[\widehat{f_{\Phi}}]^{1-\lambda}=\int_{\Rset}\left[\exp\left(\frac{\Phi(f(x))}{f(x)}\right)\right]^{\lambda-1}f(x)\,dx=\widehat{N}_{\lambda}^{[\Phi]}[f]^{1-\lambda},
	$$
	which proves Eq. \eqref{eq:phi_exp_ent2}. Doing an analogous calculation with the Shannon entropy, we find
	\begin{equation*}
		S[\widehat{f_{\Phi}}]=-\int_{\Rset}\widehat{f_{\Phi}}(y)\log\,\widehat{f_{\Phi}}(y)\,dy=-\int_{\Rset}\frac{\Phi(f(x))}{f(x)}f(x)\,dx=H_{\Phi}[f],
	\end{equation*}
completing the proof.
\end{proof}
We are now in a position to state and prove the general functional inequalities for $\Phi$-entropies, that are the main goal of this section.
\begin{theorem}\label{th:phi_ent}
	Let $\Phi$ be as in Definition \ref{def:phi_ent}. Then, for any probability density function $f$ with $H_{\Phi}[f]$ finite, the following \emph{$\Phi$-moment--entropy inequality}
	\begin{equation}\label{ineq:phi-M-E}
		\frac{\mu_{p^*,\gamma}^{[\Phi]}[f]^{\frac{1}{p^*\gamma}}}{e^{H_{\Phi}[f]}}\geqslant K^{(0)}_{p,\gamma,1}:=\big(K^{(0)}_{p,1}\big)^{\frac{1}{\gamma}},
	\end{equation}
	holds true for any $p\in\Rset$ such that $p^*>0$ and any $\gamma>0$. If, furthermore, the density $f$ and the entropic functional $\Phi$ are such that $f$ is differentiable and the transformed density $\mathfrak{E}_{\beta}[\widehat{f_{\Phi}}]$ is absolutely continuous on $\Rset$, then the following \emph{$\Phi$-Stam inequality}
	\begin{equation}\label{ineq:phi-Stam}
		\big(F_{p,\beta}^{[\Phi]}[f]\big)^{\frac{1}{p}}e^{\beta H_{\Phi}[f]}\geqslant \kappa_{p,\beta,1}^{(1)}
	\end{equation}
	or equivalently,
	\begin{equation}\label{ineq:phi-Stambis}
		\big(\phi_{p,\beta}^{[\Phi]}[f]e^{H_{\Phi}[f]}\big)^\beta\geqslant \kappa_{p,\beta,1}^{(1)}
	\end{equation}
	holds true for any $(p,\beta)\in\Rset^2$ such that $p\geq1$ and $\beta\neq0$.
\end{theorem}
{We recall here that the constants $K^{(0)}_{p,1}$, respectively $\kappa_{p,\beta,1}^{(1)}$, are the optimal constants of the corresponding moment-entropy and triparametric Stam inequality respectively, according to the notation introduced in \eqref{ineq:bip_E-M} and \eqref{ineq:trip_Stam_extended}.}
\begin{proof}
	In order to prove \eqref{ineq:phi-M-E}, we start from the moment-entropy inequality \eqref{ineq:bip_E-M} and we apply it to the transformed density $\mathfrak{E}_{\gamma}[\widehat{f_{\Phi}}]$ and with $\lambda=1$ (when the R\'enyi entropy power reduces to the Shannon entropy power). We thus obtain that
	$$
	\frac{\sigma_{p^*}[\mathfrak{E}_{\gamma}[\widehat{f_{\Phi}}]]}{e^{S[\mathfrak{E}_{\gamma}[\widehat{f_{\Phi}]}]}}\geqslant K^{(0)}_{p,1}.
	$$
	The inequality \eqref{ineq:phi-M-E} follows readily from \eqref{eq:phi_moment2} and \eqref{eq:phi_Renyi}, {taking into account that}
	$$
	\sigma_{p^*}[\mathfrak{E}_{\gamma}[\widehat{f_{\Phi}}]]=\mu_{p^*}[\mathfrak{E}_{\gamma}[\widehat{f_{\Phi}}]]^{1/p^*}=\mu_{p^*,\gamma}[\widehat{f_{\Phi}}]^{1/p^*}
	$$ 
by \eqref{eq:phi_moment_escort} and $S[\mathfrak{E}_{\gamma}[g]]=\gamma S[g]$ by Section \ref{sec:escort}, and then by raising to the power $1/\gamma$. For the proof of \eqref{ineq:phi-Stam}, we start from the triparametric Stam inequality \eqref{ineq:trip_Stam_extended} and we apply it to the transformed density $\widehat{f_{\Phi}}$ and with $\lambda=1$. Note that, in this case, the sign condition \eqref{eq:sign_cond1} is fulfilled whenever $\beta\neq0$, and $\theta(\beta,1)=\beta$, so that the required absolute continuity is that of $\mathfrak{E}_{\beta}[\widehat{f_{\Phi}}]$. We thus obtain
	$$
	\bigg(F_{p,\beta}^{\frac{1}{p\beta}}[\widehat{f_{\Phi}}]e^{S[\widehat{f_{\Phi}}]}\bigg)^{\beta}\geqslant\kappa_{p,\beta,1}^{(1)},
	$$
	and \eqref{ineq:phi-Stam} then follows from \eqref{eq:phi_Fisher2} and \eqref{eq:phi_Renyi}.
\end{proof}
Observe that a Cram\'er-Rao inequality is also valid for $p>1$, $\beta>0$ and $\gamma>0$, under the assumptions of the $\Phi$-Stam inequality, and trivially deduced by multiplying the inequalities \eqref{ineq:phi-M-E} and \eqref{ineq:phi-Stam}, the latter raised to the power $1/\beta$, and reads
\begin{equation}\label{ineq:phi-CR}
	\mu_{p^*,\gamma}^{[\Phi]}[f]^{\frac{1}{p^*\gamma}}\big(F_{p,\beta}^{[\Phi]}[f]\big)^{\frac{1}{p\beta}}\geqslant K^{(0)}_{p,\gamma,1}\bigg(\kappa_{p,\beta,1}^{(1)}\bigg)^{\frac{1}{\beta}}.
\end{equation}
Let us also notice that the previous inequalities are sharp and the minimizers can be calculated whenever the function $x\mapsto\frac{\Phi(x)}{x}$ is invertible, as assumed at the beginning of the paper. In this case, and for $p>1$, the minimizers of the inequality \eqref{ineq:phi-M-E} satisfy

$$
\mathfrak{E}_{\gamma}\big[\widehat{(f_{\rm min})_{\Phi}}\big]=g_{p,1}=a_{p,1}e^{-|y|^{p^*}}\quad\text{(up to dilations)},
$$
whence

$$
f_{\rm min}(s)=\varphi_{\gamma}(-|y(s)|^{p^*}+\log\,a_{p,1}),\qquad s(y)=\int_{0}^{y}\frac{g_{p,1}(t)}{\varphi_{\gamma}(\log g_{p,1}(t))}\,dt,
$$
where we recall that the function $\varphi$ is the inverse of $\frac{\Phi(x)}{x}$, as defined in \eqref{eq:varphi}, and $\varphi_{\gamma}(r):=\varphi(r/\gamma)$ is the inverse of $\gamma\Phi(x)/x$. For the inequality \eqref{ineq:phi-Stam}, the minimizers are given by

$$
f_{\rm min}=\varphi(\log\,\widetilde{g}_{p,\beta,1}(y(s)))=\mathfrak{Z}_{\Phi}^{-1}[\widetilde{g}_{p,\beta,1}](s),
$$
up to translations and dilations. Here we are assuming that $\log g_{p,1}$ (respectively $\log\widetilde{g}_{p,\beta,1}$) takes its values in the domain of $\varphi_{\gamma}$ (respectively $\varphi$). Notice that for the Tsallis entropies \eqref{eq:Tsallis} with $\alpha>1$ this fails, and the constant in \eqref{ineq:phi-M-E} is not attained.

\subsection{Power based $\Phi$-transformations}

We next introduce a different family of transformations, which is only well defined under the assumption that $\Phi(f)\neq0$ on the support of $f$. Note that, since $\Phi(f)$ can take the value zero at some point (for example when $f(x)=1$ in the noticeable examples recalled at the beginning of this section), the integral defining the change of variable in the definition below can be improper and we have to assume its convergence and thus restrict the range of density functions $f$ for which the following transformations are applicable.
\begin{definition}\label{def:phi_transf2}
	Let $\Phi$ and $f$ as in Definition \ref{def:phi_ent}. For any real number $k\neq0$, we introduce the \emph{$(\Phi,k)$-transformation} as follows:
	\begin{equation}\label{eq:phi_transfk}
		\mathfrak{Z}_{\Phi}^{[k]}[f](y):=\left|\frac{\Phi(f(x(y)))}{f(x(y))}\right|^{k}, \quad y(x)=\int_{x_0}^xf(s)\left|\frac{f(s)}{\Phi(f(s))}\right|^k\,ds,
	\end{equation}
	where $x_0$ is the initial point of the support of $f$, which is well defined for any density $f$ for which $y(x)<\infty$ for any $x\in(x_0,x_f)$. For simplicity, we will employ the shorter notation
	$$
	\widehat{f_{k}^{[\Phi]}}\equiv\mathfrak{Z}_{\Phi}^{[k]}[f].
	$$
\end{definition}
In a similar way as in the case of the $\Phi$-transformation, we have the following basic properties of the $(\Phi,k)$-transformation, obtained by direct calculation along the lines of the proof of Lemma \ref{lem:deriv}:
\begin{lemma}\label{lem:deriv2}
	Let $f$ be a probability density function and $\Phi$ an entropic functional as in Definition \ref{def:phi_ent} such that $\Phi(f)\neq0$ on the support of $f$. Then $\widehat{f_{k}^{[\Phi]}}$ is a probability density function as well. Moreover, if $f$ is differentiable, then $\widehat{f_{k}^{[\Phi]}}$ is also differentiable and
	\begin{equation}\label{eq:deriv_phi2}
		\frac{d\widehat{f_{k}^{[\Phi]}}(y)}{dy}=k\frac{|\Phi(f(x))|^{2k}f'(x)}{f^{2k+2}(x)}\left(\frac{\Phi'(f(x))}{\Phi(f(x))}f(x)-1\right).
	\end{equation}
\end{lemma}
Based on the previous transformations, we introduce generalized $(\Phi,k,p)$-moments and generalized $(\Phi,k,p,\beta)$-Fisher information functionals by means of the following definitions.
\begin{definition}\label{def:inf_funct2}
	Let $\Phi$ and $f$ be as in Definition \ref{def:phi_ent}. Then, for any $p\geq0$ and $\xi\neq0$, we define the $(\Phi,k,p)$-moments of $f$ by
	\begin{equation}\label{eq:phi_moment3}
		\mu_{p;k}^{[\Phi]}[f]:=\int_{\Rset}\left|\int_{x_0}^{x}\frac{f^{k+1}(t)}{|\Phi|^{k}(f(t))}\,dt\right|^pf(x)\,dx, \quad \sigma^{[\Phi]}_{p;k}[f]:=\left(\mu_{p;k}^{[\Phi]}[f]\right)^{\frac1{pk}},
	\end{equation}
	and the modified $\Phi$-entropy of order $\xi$ of $f$ by
	\begin{equation}
		\widetilde{N}_{\xi}^{[\Phi]}[f]:=\left[\int_{\Rset}f(x)\left|\frac{\Phi(f(x))}{f(x)}\right|^{\xi}\,dx\right]^{-\frac{1}{\xi}}.
	\end{equation}
	Moreover, if $f$ is differentiable, the $(\Phi,p,\beta,k)$-Fisher information of $f$ is defined as
	\begin{equation}\label{eq:phi_Fisher3}
		F_{p,\beta;k}^{[\Phi]}[f]:=k^p\int_{\Rset}f(x)^{1-(k\beta+2)p}|\Phi(f(x))|^{k\beta p}|f'(x)|^p\left|\frac{\Phi'(f(x))}{\Phi(f(x))}f(x)-1\right|^p\,dx.
	\end{equation}
We also define the following quantity:
		\begin{equation}
			\phi_{p,\beta;k}^{[\Phi]}[f]=\left(F_{p,\beta;k}^{[\Phi]}[f]\right)^{\frac1{p\beta k}}.
		\end{equation}
\end{definition}
The following result motivates the definition of the previous informational functionals.
\begin{proposition}\label{prop:funct2}
	Let $f$ and $\Phi$ be as in Definition \ref{def:phi_ent} and assume that the transformations in Definition \ref{def:phi_transf2} are applicable to $f$. Then
	\begin{equation}\label{eq:phi_moment4}
		\mu_{p;k}^{[\Phi]}[f]=\mu_p[\widehat{f_{k}^{[\Phi]}}], \quad p\geq0,
	\end{equation}
	and
	\begin{equation}\label{eq:phi_Fisher4}
		F_{p,\beta;k}^{[\Phi]}[f]=F_{p,\beta}[\widehat{f_{k}^{[\Phi]}}].
	\end{equation}
	Moreover,
	\begin{equation}\label{eq:phi_ent2}
		\widetilde{N}_{k(\lambda-1)}^{[\Phi]}[f]^k=N_{\lambda}[\widehat{f_{k}^{[\Phi]}}].
	\end{equation}
\end{proposition}
\begin{proof}
	Recalling \eqref{eq:phi_transfk}, we infer from the definition of the $p$-th moment that
	$$
	\mu_{p}[\widehat{f_{k}^{[\Phi]}}]=\int_{\Rset}|y|^p\widehat{f_{k}^{[\Phi]}}(y)\,dy
	=\int_{\Rset}\left|\int_{x_0}^{x}\frac{f^{k+1}(t)}{|\Phi|^{k}(f(t))}\,dt\right|^pf(x)\,dx
	=\mu_{p;k}^{[\Phi]}[f].
	$$
	In order to prove \eqref{eq:phi_Fisher4}, we take into account \eqref{eq:deriv_phi2} and compute
	\begin{equation*}
		\begin{split}
			F_{p,\beta}[\widehat{f_{k}^{[\Phi]}}]&=\int_{\Rset}\left[\widehat{f_{k}^{[\Phi]}}\right]^{1+(\beta-2)p}(y)\left|\frac{d}{dy}\widehat{f_{k}^{[\Phi]}}(y)\right|^p\,dy\\
			&\hspace{-2.5em}=k^p\int_{\Rset}f(x)\left|\frac{\Phi(f(x))}{f(x)}\right|^{kp(\beta-2)}\left|\frac{|\Phi|^{2k}(f(x))f'(x)}{f^{2k+2}(x)}
			\left(\frac{\Phi'(f(x))}{\Phi(f(x))}f(x)-1\right)\right|^p\,dx\\
			&\hspace{-2.5em}=k^p\int_{\Rset}f(x)^{1-(k\beta+2)p}|\Phi(f(x))|^{k\beta p}|f'(x)|^p\left|\frac{\Phi'(f(x))}{\Phi(f(x))}f(x)-1\right|^p\,dx=F_{p,\beta;k}^{[\Phi]}[f],
		\end{split}
	\end{equation*}
	as desired. Finally, we calculate, for $\lambda\neq1$,
	\begin{equation*}
		\begin{split}
		N_{\lambda}[\widehat{f_{k}^{[\Phi]}}]&=\left[\int_{\Rset}\left[\widehat{f_{k}^{[\Phi]}}(y)\right]^{\lambda}\,dy\right]^{\frac{1}{1-\lambda}}
		=\left[\int_{\Rset}\frac{|\Phi|^{k(\lambda-1)}(f(x))}{f^{k(\lambda-1)}(x)}f(x)\,dx\right]^{\frac{1}{1-\lambda}}\\&=\widetilde{N}_{k(\lambda-1)}^{[\Phi]}[f]^k,
		\end{split}
	\end{equation*}
	establishing thus \eqref{eq:phi_ent2}.
\end{proof}

\begin{remark} Letting $\lambda=1+\frac{1}{k}$ in the equality Eq. \eqref{eq:phi_ent2}, we obtain a kind of ``absolute $\Phi$-entropy", that is,
$$
\widetilde{N}_{1}^{[\Phi]}[f]^{-1}=\int_{\Rset}|\Phi(f(x))|\,dx.
$$
Observe that the right-hand side of the previous identity is not the proper $\Phi$-entropy, as the function $\Phi$ might change sign.
\end{remark}
We are now in a position to state and prove two new functional inequalities for $\Phi$-entropies. They are similar to the ones given in Theorem \ref{th:phi_ent}, but involving the moments and Fisher information defined in \eqref{eq:phi_moment3} and \eqref{eq:phi_Fisher3}.
\begin{theorem}\label{th:phi_ent2}
	Let $\Phi$ be as in Definition \ref{def:phi_ent}. Then, for any probability density function $f$ for which Definition \ref{def:phi_transf2} is applicable, the following \emph{$(\Phi,k)$-moment--entropy inequality}
		\begin{equation}\label{ineq:phi-M-E2_gen}
			\frac{\sigma_{p^*;k}^{[\Phi]}[f]}{\widetilde N_{\xi}^{[\Phi]}[f]}\geqslant \left(K^{(0)}_{p,1+\frac{\xi}{k}}\right)^{\frac1{k}},
		\end{equation}
	holds true for any $p\in\Rset$ such that $p^*>0$, for any real number $k>1$ and any $\xi\in\big(-\frac{kp^*}{1+p^*},+\infty\big)\setminus\{0\}$. Assume, furthermore, that $p\geqslant1$ and the following sign condition is fulfilled:
	\begin{equation}\label{eq:cond_sign_phi}
\sign\left(p^*\beta k+\lambda\right)=\sign(\beta k-\lambda)\neq0.
	\end{equation}
If the density $f$ and the entropic functional $\Phi$ are such that $f$ is differentiable and the transformed density $\widehat{f_{k\beta-\lambda}^{[\Phi]}}$, defined by \eqref{eq:phi_transfk} with the exponent $k\beta-\lambda$, is absolutely continuous on $\Rset$, we have the following \emph{$(\Phi,k)$-Stam inequality}
		\begin{equation}\label{ineq:phi-Stam2}
			\left[\phi_{p,\beta;k}^{[\Phi]}[f]\widetilde{N}_{\lambda}^{[\Phi]}[f]\right]^{k\beta-\lambda}\geqslant\kappa_{p,\beta,1+\frac{\lambda}{k}}^{(1)}.
		\end{equation}
\end{theorem}
\begin{proof}
	In order to prove \eqref{ineq:phi-M-E2_gen}, we start from the moment-entropy inequality \eqref{ineq:bip_E-M} and we apply it to the transformed density $\widehat{f_{k}^{[\Phi]}}$ with $\lambda=1+\xi/k$, which satisfies \eqref{eq:param_clas} by the assumption on $\xi$, and
	$$
		\frac{\sigma_{p^*}[\widehat{f_{k}^{[\Phi]}}]}{N_{\lambda}[\widehat{f_{k}^{[\Phi]}}]}\geqslant K^{(0)}_{p,\lambda},
		$$
		and the inequality \eqref{ineq:phi-M-E2_gen} follows readily from \eqref{eq:phi_moment4}, \eqref{eq:phi_moment3} and \eqref{eq:phi_ent2}, since $k(\lambda-1)=\xi$.
For the proof of \eqref{ineq:phi-Stam2}, we start from the triparametric Stam inequality \eqref{ineq:trip_Stam_extended} and we apply it to the transformed density $\widehat{f_{k}^{[\Phi]}}$. Note that, for every $\theta\neq0$,
\[
\big(\widehat{f_{k}^{[\Phi]}}\big)^{\theta}=\Big|\frac{\Phi(f)}{f}\Big|^{k\theta}\qquad\text{and}\qquad
\big(\widehat{f_{k}^{[\Phi]}}\big)^{1-\theta}dy=f\,\Big|\frac{\Phi(f)}{f}\Big|^{-k\theta}dx ,
\]%
so that, by \eqref{eq:escort} and \eqref{eq:phi_transfk},
\[
\mathfrak{E}_{\theta}\big[\widehat{f_{k}^{[\Phi]}}\big]=\widehat{f_{k\theta}^{[\Phi]}}\qquad\text{up to a translation,}
\]%
and the required absolute continuity is that of $\widehat{f_{k(1+\beta-\lambda)}^{[\Phi]}}$. We thus obtain
	$$
		\left(F_{p,\beta}^{\frac{1}{p\beta}}[\widehat{f_{k}^{[\Phi]}}]N_{\lambda}[\widehat{f_{k}^{[\Phi]}}]\right)^{1+\beta-\lambda}
		\geqslant\kappa_{p,\beta,\lambda}^{(1)}
		$$
whenever the condition~\eqref{eq:sign_cond1} is fulfilled. From \eqref{eq:phi_Fisher4} and \eqref{eq:phi_ent2} follows that
		$$
		\left(\phi_{p,\beta;k}^{[\Phi]}[f]\widetilde N_{k(\lambda-1)}^{[\Phi]}[f]\right)^{k(1+\beta-\lambda)}
		\geqslant\kappa_{p,\beta,\lambda}^{(1)}.
		$$
Adopting the notation $\overline{\lambda}:=k(\lambda-1)$, it is an immediate verification that the condition \eqref{eq:sign_cond1} is transformed into Eq. \eqref{eq:cond_sign_phi} after dropping the overline. Moreover,
$$
k(1+\beta-\lambda)=k\beta-\overline{\lambda}, \quad \lambda=1+\frac{\overline\lambda}{k},
$$
and we arrive at the inequality \eqref{ineq:phi-Stam2} after easy manipulations and dropping the overline from the notation; in particular, the absolute continuity assumption becomes that of $\widehat{f_{k\beta-\lambda}^{[\Phi]}}$.
\end{proof}

\section{Applications to the Tsallis, Kaniadakis and general group entropies}\label{sec:applications}

The inequalities given in Theorem \ref{th:phi_ent} apply, in particular, to a number of interesting entropies, such as the Tsallis entropies corresponding to $\Phi$ given in \eqref{eq:Tsallis} (the $\Phi$-Stam inequality only for $\alpha\in(0,1)$, since for $\alpha>1$ its assumption is never fulfilled), the Kaniadakis entropies defined for the particular expression of $\Phi$ given in \eqref{eq:Kania}, and general group entropies according to the axioms and constructions introduced in Section \ref{sec:group-axioms}.

\subsection{The Tsallis entropy}\label{sec:Tsallis}

Let us recall that the Tsallies entropy (or family of entropies depending on the parameter $\alpha$) is the only composable entropy in trace form, as established in \cite{Enciso2017}. It is thus expected to have some better properties than the rest of the family of group entropies we deal with, in particular with respect to the scaling invariance. Indeed, in an abstract framework, due to the presence of a general function $\Phi$, which might be either not homogeneous at all with respect to a scaling change, or its terms to have different homogeneity (as it happens in the case of the Kaniadakis entropies and the more general group entropies considered in Section \ref{sec:group}), the informational functionals introduced in Section \ref{sec:ineq} are not expected to behave well with respect to scaling changes of a density $f$. However, we shall see that the Tsallis entropies actually present rather nice properties with respect to a scaling. Denoting by
\begin{equation}\label{eq:scaling}
f^{[a]}(x):=af(ax), \quad a>0,
\end{equation}
we start from computing, for $\alpha\neq1$,
\begin{equation}\label{eq:interm3}
\begin{split}
e^{T_{\alpha}[f^{[a]}]}&=\exp\left[\frac{1}{\alpha-1}\left(1-\int_{\Rset}a^{\alpha}f^{\alpha}(ax)\,dx\right)\right]\\
&=\exp\left[\frac{1}{\alpha-1}\left(1-a^{\alpha-1}\int_{\Rset}f^{\alpha}(x)\,dx\right)\right]\\
&=\exp\left(-\frac{a^{\alpha-1}-1}{\alpha-1}\right)\exp\left[\frac{a^{\alpha-1}}{\alpha-1}\left(1-\int_{\Rset}f^{\alpha}(x)\,dx\right)\right]\\
&=K(a,\alpha)^{-1}e^{a^{\alpha-1}T_{\alpha}[f]},
\end{split}
\end{equation}
where
$$
K(a,\alpha):=\exp\left(\frac{a^{\alpha-1}-1}{\alpha-1}\right)=\exp(\log_{G}(a)),
$$
the latter equality being expressed in terms of the group logarithm associated to the Tsallis entropy of order $\alpha$,
$$
\log_{G}(t)=\frac{t^{\alpha-1}-1}{\alpha-1}.
$$
We next particularize the general definition of the $(\Phi,p,\beta)$-Fisher information given in \eqref{eq:phi_Fisher} to the Tsallis entropy.
\begin{equation}\label{eq:interm4}
\begin{split}
F_{p,\beta}^{(T)}[f]&=\int_{\Rset}f^{1-p}(x)e^{\frac{p\beta(f^{\alpha-1}(x)-1)}{\alpha-1}}\left|\frac{d}{dx}\left(\frac{f^{\alpha-1}(x)-1}{\alpha-1}\right)\right|^p\,dx\\
&=\int_{\Rset}f^{1+p(\alpha-3)}(x)|f'(x)|^pe^{\frac{p\beta(f^{\alpha-1}(x)-1)}{\alpha-1}}\,dx.
\end{split}
\end{equation}
Gathering \eqref{eq:interm3} and \eqref{eq:interm4}, we calculate the $(\Phi,p,\beta)$-Fisher information corresponding to the Tsallis entropy of a scaled density $f^{[a]}$ as follows:
\begin{equation}\label{eq:interm5}
\begin{split}
F_{p,\beta}^{(T)}[f^{[a]}]&=\int_{\Rset}a^{1+p(\alpha-3)}f^{1+p(\alpha-3)}(ax)a^{2p}|f'(ax)|^p\\
&\quad\quad\times\exp\left[\frac{p\beta}{\alpha-1}\left(a^{\alpha-1}f^{\alpha-1}(ax)-1\right)\right]\,dx\\
&=a^{p(\alpha-1)}\int_{\Rset}f^{1+p(\alpha-3)}(x)|f'(x)|^p\exp\left[\frac{p\beta a^{\alpha-1}}{\alpha-1}\left(f^{\alpha-1}(x)-1\right)\right]\\
&\quad\quad\times\exp\left[\frac{p\beta(a^{\alpha-1}-1)}{\alpha-1}\right]\,dx\\
&=a^{p(\alpha-1)}K(a,\alpha)^{p\beta}\int_{\Rset}f^{1+p(\alpha-3)}(x)|f'(x)|^p\\
&\quad\quad\times\exp\left[\frac{p\beta a^{\alpha-1}}{\alpha-1}\left(f^{\alpha-1}(x)-1\right)\right]\,dx
\\
&=a^{p(\alpha-1)}K(a,\alpha)^{p\beta}F_{p,a^{\alpha-1}\beta}^{(T)}[f].
\end{split}
\end{equation}
We deduce from \eqref{eq:interm5} that the $\Phi$-Stam inequality \eqref{ineq:phi-Stam} applied to $f^{[a]}$ (for $\alpha\in(0,1)$ and $f^{[a]}$ satisfying the assumption of Theorem \ref{th:phi_ent}) reads
$$
\left(F_{p,a^{\alpha-1}\beta}^{(T)}[f]\right)^{\frac{1}{p}}e^{a^{\alpha-1}\beta T_{\alpha}[f]}\geqslant a^{1-\alpha}\kappa_{p,\beta,1}^{(1)}=\kappa_{p,a^{\alpha-1}\beta,1}^{(1)},
$$
where the last equality follows from $\kappa_{p,\beta,1}^{(1)}=\kappa_{p,1,1}^{(1)}/|\beta|$, see \cite[Section 5]{IP2025}, or equivalently (for $\beta>0$)
$$
	\phi_{p,a^{\alpha-1}\beta}^{(T)}[f]e^{T_{\alpha}[f]}\geqslant\left[\kappa_{p,a^{\alpha-1}\beta,1}^{(1)}\right]^\frac{a^{1-\alpha}}{\beta}.
$$
{The latter expressions show} that the inequality \eqref{ineq:phi-Stam} for $f^{[a]}$ with parameter $\beta$ is the inequality \eqref{ineq:phi-Stam} for $f$ with parameter $a^{\alpha-1}\beta$: the family of inequalities \eqref{ineq:phi-Stam} is invariant with respect to scaling changes, at least in the case of the Tsallis entropy. In a similar way, we note that the $(\Phi,p,\gamma)$-moment reduces in the case of the Tsallis entropy to
$$
\mu_{p,\gamma}^{(T)}[f]=\int_{\Rset}\left|\int_{x_0}^{x}\exp\left(-\gamma\frac{f^{\alpha-1}(t)-1}{\alpha-1}\right)f(t)\,dt\right|^p\,f(x)\,dx,
$$
whence
\begin{equation*}
\begin{split}
\mu_{p,\gamma}^{(T)}[f^{[a]}]&=\int_{\Rset}\left|\int_{x_0}^{x}\exp\left(-\gamma\frac{a^{\alpha-1}f^{\alpha-1}(at)-1}{\alpha-1}\right)af(at)\,dt\right|^paf(ax)\,dx\\
&=K(a,\alpha)^{-p\gamma}\int_{\Rset}\left|\int_{ax_0}^{x}\exp\left(-a^{\alpha-1}\gamma\frac{f^{\alpha-1}(t)-1}{\alpha-1}\right)f(t)\,dt\right|^pf(x)\,dx\\
&=K(a,\alpha)^{-p\gamma}\mu_{p,\gamma a^{\alpha-1}}^{(T)}[f].
\end{split}
\end{equation*}
where, in the last line, the $\Phi$-moment of $f$ is taken with respect to the base point $ax_0$. By inserting the previous equality and \eqref{eq:interm3} into the $\Phi$-moment-entropy inequality, we observe that the constant $K(a,\alpha)$ simplifies from the final expression and we get a similar inequality for $f^{[a]}$.

\subsection{The Kaniadakis entropy}

In the case of the Kaniadakis entropy, some of the previous general calculations and inequalities can be expressed in terms of hyperbolic functions. More precisely, starting from \eqref{eq:Kania}, we obtain
$$
\frac{\Phi(t)}{t}=\frac{t^{\alpha}-t^{-\alpha}}{2\alpha}=\frac{1}{\alpha}\sinh(\alpha\log\,t)=:\log_{K}(t),
$$
or, equivalently,
\begin{equation}\label{eq:Kania2}
\Phi(t)=\frac{t}{\alpha}\sinh(\alpha\log\,t).
\end{equation}
We can thus differentiate in \eqref{eq:Kania2} to find
$$
\Phi'(t)=\frac{1}{\alpha}\sinh(\alpha\log\,t)+\cosh(\alpha\log\,t),
$$
hence
$$
\frac{\Phi'(t)}{\Phi(t)}=\frac{1}{t}+\frac{\alpha}{t}\coth(\alpha\log\,t).
$$
The latter equality implies
\begin{equation}\label{eq:interm2}
\frac{\Phi'(f(x))}{\Phi(f(x))}f(x)-1=\alpha\coth(\alpha\log\,f(x)).
\end{equation}
We proceed with the calculation of the $(\Phi,p,\beta)$-Fisher information, that we denote, in order to particularize to the Kaniadakis entropy, by $F_{p,\beta}^{(K)}$. We infer from \eqref{eq:phi_Fisher}, \eqref{eq:interm2} and \eqref{eq:Kania2} that
\begin{equation*}
\begin{split}
F_{p,\beta}^{(K)}[f]&
=\int_{\Rset}f(x)^{1-p}\exp\left(\frac{p\beta\Phi(f(x))}{f(x)}\right)\left|\frac{d}{dx}\left(\frac{\Phi(f(x))}{f(x)}\right)\right|^p\,dx\\
&=\int_{\Rset}f(x)^{1-p}\exp\left(\frac{p\beta}{\alpha}\sinh(\alpha\log\,f(x))\right)\\
&\qquad \qquad \times\left|\frac{\Phi'(f(x))f'(x)f(x)-\Phi(f(x))f'(x)}{f^2(x)}\right|^p\,dx\\
&=\frac{1}{|\alpha|^p}\int_{\Rset}f(x)^{1-2p}|\sinh(\alpha\log\,f(x))|^p\exp\left(\frac{p\beta}{\alpha}\sinh(\alpha\log\,f(x))\right)\\
&\qquad \qquad \times\left|\alpha\coth(\alpha\log\,f(x))\right|^p|f'(x)|^p\,dx\\
&=\int_{\Rset}f(x)^{1-2p}\exp\left(\frac{p\beta}{\alpha}\sinh(\alpha\log\,f(x))\right)\cosh(\alpha\log\,f(x))^p|f'(x)|^p\,dx.
\end{split}
\end{equation*}
We go on with the calculation of the $(\Phi,p)$-moments $\mu_{p}^{(K)}$. We have
\begin{equation*}
\begin{split}
\mu_{p}^{(K)}[f]&=\int_{\Rset}\left|\int_{x_0}^{x}\exp\left(-\frac{\Phi(f(t))}{f(t)}\right)f(t)\,dt\right|^pf(x)\,dx\\
&=\int_{\Rset}\left|\int_{x_0}^{x}\exp\left(-\frac{1}{\alpha}\sinh(\alpha\log\,f(t))\right)f(t)\,dt\right|^pf(x)\,dx.
\end{split}
\end{equation*}
Thus, the inequalities \eqref{ineq:phi-M-E}, \eqref{ineq:phi-Stam} and (when it holds true) \eqref{ineq:phi-CR} can be written, in the important particular case of the Kaniadakis entropy, in terms of the more explicit previous informational functionals.

\subsection{General group entropies}\label{sec:group}

Let $\Phi:[0,\infty) \to \mathbb{R}$ be a convex function such that $\Phi(0)=0$ and $\Phi\in C^{2}(0,\infty)$, and assume moreover that $\Phi(1)=0$, $\Phi'(1)=1$ and $t^{2}\Phi''(t)<2\big(t\Phi'(t)-\Phi(t)\big)$ for $t>0$. {The latter conditions ensure that} $\Phi(t)/t$ is strictly increasing and strictly concave. To any such function, we can associate a group-theoretical logarithm, in the sense of Section \ref{sec:23}:
$$
\Phi(t)= t G(\ln t) = t \log_{G}(t),
$$
or, equivalently
$$
G(t)= \frac{\Phi(\exp t)}{\exp t}.
$$
We can thus associate a formal group law by defining
\begin{equation*}
\Psi(x,y)= G(G^{-1}(x)+G^{-1}(y)),
\end{equation*}
showing thus that any such $\Phi$-entropy can be approached by employing the previous theoretical framework. In particular, for the inverse function of $\frac{\Phi(t)}{t}$ we have
$$
\varphi^{-1}(t)=\frac{\Phi(t)}{t}= \log_{G}(t), \quad {\rm that \ is,} \quad \varphi(t)= \exp_{G}(t).
$$
Taking into account the previous equalities, one can further deduce that
$$
\Psi(x,y)= \frac{\Phi(\exp(\ln(\varphi(x))+\ln (\varphi(y))))}{\exp(\ln(\varphi(x))+\ln (\varphi(y)))}= \frac{\Phi(\varphi(x)\varphi(y))}{\varphi(x)\varphi(y)}
$$
Recalling the transformations introduced in \eqref{eq:phi_transf} and \eqref{eq:phi_transfk}, we observe that they can be interpreted in terms of the group logarithm introduced above. More precisely, for any probability density function $f$ and any $k>1$ we have
\begin{equation}\label{eq:corresp}
\mathfrak{Z}_{\Phi}[f](y)= \exp(\log_{G}(f(x(y)))), \quad \mathfrak{Z}_{\Phi}^{[k]}[f](y)=|\log_{G}(f(x(y)))|^k,
\end{equation}
while the change of variables is also defined via the group logarithm for both transformations:
$$
y'(x)= f(x)\exp(-\log_{G}(f(x))), \quad {\rm respectively} \quad y'(x)=f(x)|\log_{G}(f(x))|^{-k}.
$$
Taking into account \eqref{eq:corresp}, the functionals introduced in Definition \ref{def:inf_funct} can be expressed in terms of the group logarithm as well. Thus, \eqref{eq:phi_Fisher} reads
\begin{equation}\label{eq:group-Fisher}
F_{p,\beta}^{(G)}[f]=\int_{\Rset}f^{1-p}(x)\exp(p\beta\log_{G}\,f(x))\left|(\log_{G} (f(x))) '\right|^p\,dx,
\end{equation}
while \eqref{eq:phi_moment} is equivalent to
\begin{equation}\label{eq:group-moment}
\mu_{p}^{(G)}[f]=\int_{\Rset}\left|\int_{x_0}^x\exp\big[-\log_{G}(f(t))\big]f(t)\,dt\right|^pf(x)\,dx.
\end{equation}
Moreover, the $\Phi$-entropy itself can be written as
\begin{equation}\label{eq:group-ent}
H_{\Phi}^{(G)}[f]=-\int_{\Rset}\frac{\Phi(f(x))}{f(x)}\,f(x)\,dx=-\int_{\Rset}f(x)\log_{G}\,f(x)\,dx,
\end{equation}
generalizing thus the usual expression of the Shannon entropy to the generalized logarithms. It follows that the inequalities \eqref{ineq:phi-M-E} and \eqref{ineq:phi-Stam} can be written in terms of the generalized logarithms by substituting the $(\Phi,p)$-moments, the $(\Phi,p,\beta)$-Fisher information and the $\Phi$-entropy itself by the expressions given in \eqref{eq:group-moment}, \eqref{eq:group-Fisher} and \eqref{eq:group-ent}, respectively. 

We end this section by examining the behavior of the previous informational functionals with respect to a scaling change. Recalling the notation $f^{[a]}$ introduced in \eqref{eq:scaling}, we have 
\begin{equation*}
\begin{split}
H_{\Phi}^{(G)}[f^{[a]}]&=-\int_{\Rset}af(ax)\log_{G}\,af(ax)\,dx=-\int_{\Rset}f(x)\log_{G}(af(x))\,dx\\
&=-\int_{\Rset}f(x)G(\ln\,a+\ln\,f(x))\,dx
\end{split}
\end{equation*}
and 
\begin{equation*}
\begin{split}
\mu_{p}^{(G)}[f^{[a]}]&=\int_{\Rset}\left|\int_{x_0}^x\exp\big[-\log_{G}(af(at))\big]af(at)\,dt\right|^paf(ax)\,dx\\
&=\int_{\Rset}\left|\int_{ax_0}^{ax}\exp\big[-\log_{G}(af(y))\big]f(y)\,dy\right|^paf(ax)\,dx\\
&=\int_{\Rset}\left|\int_{ax_0}^x\exp\big[-G(\ln\,a+\ln\,f(y))\big]f(y)\,dy\right|^pf(x)\,dx.
\end{split}
\end{equation*}
Finally, for the $(\Phi,p,\beta)$-Fisher information we have
\begin{equation*}
\begin{split}
F_{p,\beta}^{(G)}[f^{[a]}]&=\int_{\Rset}(af(ax))^{1-p}\exp(p\beta\log_{G}(af(ax)))\left|\frac{d}{dx}\log_{G}(af(ax))\right|^p\,dx\\
&=\int_{\Rset}f(x)^{1-p}\exp(p\beta\log_{G}(af(x)))\left|\frac{d}{dx}\log_{G}(af(x))\right|^p\,dx\\
&=\int_{\Rset}f(x)^{1-p}\exp(p\beta G(\ln\,a+\ln\,f(x)))\left|\frac{d}{dx}G(\ln\,a+\ln\,f(x))\right|^p\,dx,
\end{split}
\end{equation*}
where we recall in all the previous expressions that $G(t)=\frac{\Phi(e^t)}{e^t}$. The previous expressions represent the generalizations to general group entropies of the scaling invariance identified for the Tsallis entropy in Section \ref{sec:Tsallis}, although in the general case the expressions do not have such an elegant form as in the case of the Tsallis entropy. In general, the left-hand sides of \eqref{ineq:phi-M-E} and \eqref{ineq:phi-Stam} are not invariant under scaling changes.

\subsection*{Acknowledgements}

R. G. I. is partially supported by the project PID2024-160967NB-I00 (AEI) funded by the Ministry of Science, Innovation and Universities of Spain and FEDER (EU). D. P.-C. is partially supported by the project PID2023-153035NB-I00 (AEI) funded by the Ministry of Science, Innovation and Universities of Spain and “ERDF/EU A way of making Europe”. D. P.-C. also thanks Universidad Rey Juan Carlos for funding through the project 2025/SOLCON-160677. 
The research of P. T. has been partly supported by the Project PID2024-156610NB-I00 of Ministerio de Ciencias, Innovaci\'on y Universidades,  and by  the Severo Ochoa Programme for Centres of Excellence in R\&D
(CEX2019-000904-S), Ministerio de Ciencia, Innovaci\'{o}n y Universidades y Agencia Estatal de Investigaci\'on, Spain.  P.T. is a member of the Gruppo Nazionale di Fisica Matematica (GNFM) of the Istituto Nazionale di Alta Matematica (INdAM). 
\bigskip

\noindent \textbf{Data availability} Our manuscript has no associated data.

\bigskip

\noindent \textbf{Competing interest} The authors declare that there is no competing interest.

\bibliographystyle{plain}
\bibliography{refs2}
\end{document}